\documentclass[aps,prd,amsmath,amssymb,preprintnumbers,onecolumn,11pt,nofootinbib]{revtex4}
\usepackage[a4paper,left=20mm,right=20mm,top=25mm,bottom=25mm]{geometry}
\usepackage{amsmath,amssymb}
\usepackage{mathrsfs}
\usepackage{graphicx}
\usepackage{color}
\usepackage{subfigure}
\usepackage{fancyhdr}
\usepackage{multirow}
\usepackage{float}
\usepackage{epsfig}
\usepackage{amsfonts}
\usepackage{bm}
\usepackage{xcolor}

\definecolor{CP3}{cmyk}{0,0.88,0.77,0.40}

\newcommand{\be}{\begin{equation}}
\newcommand{\ee}{\end{equation}}
\newcommand{\ba}{\begin{eqnarray}}
\newcommand{\ea}{\end{eqnarray}}

\renewcommand{\(}{\left(}
\renewcommand{\)}{\right)}
\renewcommand{\[}{\left[}
\renewcommand{\]}{\right]}

\newcommand\Mp{m_{\rm pl}}
\newcommand\barm{\bar{m}_\phi}
\newcommand\tN{\tilde{N}}

\begin{document}

\title{Observational Constraints and Preheating in Cuscuton Inflation}

\author{Phongpichit Channuie}
\email{channuie@gmail.com}
\affiliation{College of Graduate Studies, Walailak University, Thasala, Nakhon Si Thammarat, 80160, Thailand}
\affiliation{School of Science, Walailak University, Thasala, Nakhon Si Thammarat, 80160, Thailand}

\author{Khamphee Karwan}
\email{khampheek@nu.ac.th}
\affiliation{The Institute for Fundamental Study “The Tah Poe Academia Institute”, Naresuan University, Phitsanulok 65000, Thailand}

\author{Jakkrit Sangtawee}%
\email{jakkrits60@nu.ac.th}
\affiliation{The Institute for Fundamental Study “The Tah Poe Academia Institute”, Naresuan University, Phitsanulok 65000, Thailand}

\date{\today}

\begin{abstract}

We study cuscuton inflation for the models where the potential of the cuscuton takes quadratic and exponential forms. We find that for the quadratic potential, a scalar spectral index $n_s$ is not affected by cuscuton at the leading order in the slow-roll inflation models. However, a tensor-to-scalar ratio $r$ can be suppressed.
For the exponential potential of cuscuton, we find the condition for which the inflation has a graceful exit. 
Under this condition, the observational predictions in this model differ by a few percent from those found in standard inflation. 
We also examine the particle production due to parametric resonances in both models. We find that in Minkowski space the stage of parametric resonances can be described by the Mathieu equation. 
For the case where the cuscuton has quadratic potential, the amplitude of the driving force in the Mathieu equation has a similar form as that in standard inflation. Nevertheless, in the case of exponential potential, the amplitude of the driving force decreases faster than that in the standard case.
However, parametric resonances in our models can be sufficiently broad possible for the exponential growth of the number of particles.
We briefly discuss the case in which the expansion of space is taken into account.

\end{abstract}


\maketitle

\section{Introduction}

A successful scenario for solving shortcomings of the hot Big Bang model and providing a mechanism for creating the primordial density perturbation is based on the accelerated expansion of the early universe driven by a scalar field known as inflaton.
This scenario is the cosmic inflation \cite{Starobinsky:1980te,Guth:1980zm,Linde:1981mu, Albrecht:1982wi}. Notice that the pioneer work \cite{Starobinsky:1980te}, where the first full and internally self-consistent inflationary model with the graceful exit from inflation to the final radiation dominated state was developed in the $R+R^2$ modified gravity, which remains viable by now. In principle, the cosmic inflation can be achieved by introducing an additional degree of freedom, e.g., inflaton, in the universe.
Hence, the cosmic inflation can also be realized in modified theories of gravity such as the scalar-tensor theories of gravity or $f(R)$ gravity \cite{Starobinsky:1980te} due to the existence of a scalar degree of freedom for gravity in these theories, (see also review articles, e.g., \cite{Sotiriou:2008rp,DeFelice:2010aj}).

Usually, the modification of the theories of gravity can be done by adding extra degrees of freedom to gravitational interaction in the General relativity (GR) \cite{Fujii2009}.
However, there are many attempts to construct a class of minimally modified theories of gravity in which the degrees of freedom for gravity are two tensorial modes similar to GR \cite{cov, Mukohyama:17, Mukohyama:19, DeFelice:2020eju, Gao:20}. 
This class of theories could be constructed by breaking the temporal diffeomorphism invariance of GR.
Cosmology in these theories has been investigated in \cite{Aoki:2018brq, Aoki:20, jk}.
A possible form of  minimally modified theories of gravity is the cuscuton theory \cite{Afshordi:2006ad, Afshordi:2007yx} which is constructed by adding a minimally coupled scalar field to GR.
To keep the total degrees of freedom unchanged,
the scalar field namely the cuscuton field has to be a non-dynamical field in the unitary gauge.
However, the existence  of the cuscuton field alters the constraint equation of GR and the dispersion relation of the field coupled to the cuscuton.

Influences of cuscuton on Cosmic Microwave Background (CMB) have been studied in \cite{Afshordi:2007yx}.
Bouncing  cosmology has been investigated in \cite{Boruah:2018pvq,Quintin:2019orx,Kim:2020iwq}. 
In \cite{Ito:2019fie}, it has been shown that power-law solutions of inflation  become viable in cuscuton gravity.
A solution describing an accelerating universe with a stable extra dimension in cuscuton gravity has been studied in  \cite{Ito:2019ztb}.
Observable predictions in Generalized cuscuton models of inflation have been analyzed in \cite{Bartolo:2021wpt}. In this work, we revisit the model proposed in \cite{Bartolo:2021wpt} and estimate observational predictions of the cuscuton inflation with two particular forms of the cuscuton potential. We also study preheating effects in the cuscuton inflation. 

Regarding existing literature, the preheating mechanism was examined in \cite{Greene:1997fu,Kaiser:1995fb,Son:1996uv}. The properties of resonance of models with non-minimally coupled scalar field ${\chi}$ were carried out in \cite{Tsujikawa:1999jh}. The inclusion of a non-minimal coupling $\xi R\chi^{2}$ term with a sizeable range of parameter $\xi$ produced a sufficient resonance. Higher-curvature inflation models with $(R+\alpha^{n}R^{n})$ allowing to study a parametric preheating of a scalar field coupled non-minimally to a spacetime curvature were also investigated \cite{Tsujikawa:1999iv}. In \cite{vandeBruck:2016leo}, the authors studied preheating effects in the extended Starobinsky model \cite{Starobinsky:1980te} with an additional scalar field which interacts directly with the inflaton field via a four-leg interaction term. Interestingly, regarding multi-field inflation, a preheating mechanism in asymmetric $\alpha$-attractors has been discussed in \cite{Iarygina:2020dwe}, see also \cite{Nguyen:2019kbm} for preheating after multifield inflation; while Palatini formalism of gravitational dark matter production during preheating stage was carried out in \cite{Karam:2020rpa}. The study of preheating in the Palatini formalism with a quadratic inflaton potential with a $\alpha\,R^2$ term has been discussed in \cite{Karam:2021sno}.

In Sec.~(\ref{sec2}), we present our formalism of cuscuton inflation by considering two types of the cuscuton potentials. In Sec.(\ref{sec3}), we constrain the model parameters using observational data. We also investigate the preheating effect in the cuscuton inflation in Sec.~(\ref{sec4}). Here we demonstrate if parametric resonances in our models are sufficiently broad possible for the exponential growth of the number of particles and discuss the case in which the expansion of space is considered. We conclude our findings in Sec.~(\ref{sec5}).

\section{Cuscuton inflation}
\label{sec2}

We study the cuscuton inflation described by the action
\ba
S &=& \int d^4x\,\sqrt{-g} \Bigg(
\frac{\Mp^2}{2} R + \mu^2 \sqrt{- \partial_\mu \varphi \partial^\mu \varphi} - V(\varphi) 
- \frac{1}{2} \partial_\mu \phi \partial^\mu \phi - U(\phi)
\Bigg),\label{ac0}
\ea
where $\Mp = 1/ \sqrt{8\pi G}$ is the reduced Planck mass,
$\mu$ is a constant with dimension of mass,
 $V$ and $U$ are the potentials of the cuscuton field $\varphi$ and inflaton field$\phi$.

The cuscuton can be a non-dynamical field and consequently a theory of gravity has two degrees of freedom at non-linear level if the cuscuton is homogenous \cite{Gomes:2017tzd}.
Nevertheless, if the cuscuton can have inhomogeneous part, there will be instantaneous modes which enhance the degrees of freedom of gravity \cite{Bartolo:2021wpt}.
Hence, in the following consideration, the cuscuton is supposed to be a homogeneous field.

The spacetime of the background universe is described by the spatially flat Friedmann-Lema\^itre-Robertson-Walker (FLRW) metric,
\begin{align}
    ds^2 = - dt^2 + a^2(t) \delta_{ij} dx^i dx^j\,,
\end{align}
where $a(t)$ is the cosmic scale factor. 
The homogeneity and isotropy of the background universe require that the field  $\phi$ depend only on time,
so that the equations of motion are \cite{Ito:2019fie}
\begin{align}
3 \Mp^2 H^2 = V + U + \frac{1}{2} \dot\phi^2\,, 
\label{h2}\\
    \mathrm{sign}(\dot \varphi)\, 3 \mu^2 H + V_\varphi = 0\,,
\label{kgcus} \\
    \ddot \phi + 3 H \dot \phi + U_\phi =0\,,
\label{kgphi}
\end{align}
where $H=\dot a/a$ is the Hubble parameter.
During inflation, the inflaton slowly evolves so that
Eqs.~(\ref{kgphi}) and (\ref{h2}) become
\be
\frac{d \phi}{d N} \simeq  \frac{U_\phi}{3H^2}\,,
\quad\mbox{and}\quad
3 \Mp^2 H^2 \simeq V + U\,, 
\label{slow-power}
\ee 
where $N \equiv \ln(a_e/a)$ is the number of e-folding measured from the end of inflation at scale factor $a = a_e$.
The first equation in Eq.~(\ref{slow-power}) can be integrated as
\be
N_* = \int_{\phi_e}^{\phi_*}  \frac{3 H^2(\phi)}{U_{\phi}} d \phi\,,
\label{e-folding}
\ee
where subscripts ${}_e$ and ${}_*$ denote evaluation at the end of inflation and at the time when the perturbation  modes relevant to CMB anisotropy cross outside the Hubble horizon.
The Hubble parameter in Eq.~(\ref{e-folding}) can be expressed in terms of the inflaton field if the form of the cuscuton potential $V$ is specified.
According to the analysis in \cite{Afshordi:2006ad}, the cuscuton field affects dynamics of the background universe by modifying the Friedmann equation.
It follows from Eq.~(\ref{h2}) that this modification depends on the form of the potential $V$ of the cuscuton.
We are interested in the quadratic and exponential forms of the cuscuton potential,
because the cuscuton has a scaling solution when its potential takes the quadratic form, while for the exponential potential, the modified Friedmann equation takes the form as in the Dvali-Gabadadze-Poratti (DGP) brane-world model \cite{Afshordi:2007yx}.
In the following consideration,
we choose the inflaton potential in the form
\be
U(\phi) = \frac 12 m_\phi^2 \phi^2\,,
\label{inf-pot}
\ee
where the inflaton mass $m_\phi$ is constant.

The scalar spectral index $n_s$ and the tensor-to-scalar ratio $r$ in the cuscuton inflation are \cite{Bartolo:2021wpt}
\be
n_s -1 = -2 \epsilon  - \beta\,,
\quad\mbox{and}\quad
r = 16 \alpha\,,
\label{predict}
\ee
where
\begin{align}
\epsilon= - \frac{\dot H}{H^2}\,, \quad 
\alpha = \frac{\dot \phi^2}{2 \Mp^2 H^2}\,, \quad  
\beta= \frac{\dot \alpha}{\alpha H}\,.
\label{slowroll-p}
\end{align}
Note that the prediction from standard inflationary model is $r = 16\epsilon$ rather than that in Eq.~(\ref{predict}).

\section{Observational constraints}
\label{sec3}

\subsection{Quadratic potential}\label{qp}

We first choose the quadratic potential for the cuscuton in the form
\be
  V(\varphi) = \frac{1}{2}m^{2}\varphi^2(t)\,,
\label{quad}
\ee
where the mass $m$ is constant.
Substituting Eq.~(\ref{quad}) into Eq.~(\ref{kgcus}),
we get
\be
9 \mu^4 H^2 = m^{4}\varphi^2\,.
\label{vh}
\ee
Inserting Eq.~(\ref{vh}) into Eq.~(\ref{h2}),
the friedmann equation becomes
\be
\(3 \Mp^2 - \frac 92 \Lambda\) H^2 = U + \frac{1}{2} \dot\phi^2\,, 
\label{h2-pow}
\ee
where $\Lambda \equiv \mu^4 / m^2$.
To estimate observable quantities,
we compute a slow-roll parameter $\epsilon$ by differentiating Eq.~(\ref{h2-pow}) with respect to time and applying the slow-roll approximation as
\be
\epsilon \simeq \(\Mp^2 - \frac 32 \Lambda\) \frac{U_\phi^2}{2 U^2}\,, 
\label{epsilon-pow-gen}
\ee
where the slow-roll approximation $U \gg \dot\phi^2$ is used.
The number of e-folding can be computed from Eq.~(\ref{slow-power}) as
\ba
N_* &=&
 \int_{\phi_e}^{\phi_*}  \frac{1}{\Mp^2 - 3 \Lambda / 2} \frac{U}{U_\phi} d \phi
\nonumber\\
&=&  \frac{1}{2\(\Mp^2 - 3 \Lambda / 2\)} \[
\left. \frac{U^2}{U_\phi^2}\right |_{\phi_e}^{\phi_*} 
+ \int_{\phi_e}^{\phi_*} \frac{U^3}{U_\phi^3}\frac{U_{\phi\phi}}{U} d\phi
\]
\nonumber\\
&\sim&  \frac{1}{2\(\Mp^2 - 3 \Lambda / 2\)} \frac{U^2}{U_\phi^2} - \tN\,,
\label{nfold-pow-gen}
 \ea
where $\tN$ is a constant and only the dominant term is keeped in the last line.
Here, we use the approximation $U/ (\Mp U_{\phi}) \sim {\cal O}(1/ \sqrt{\epsilon})$ and $\Mp^2 U_{\phi\phi} /  U \sim {\cal O}(\epsilon)$ as well as $\epsilon \ll 1$.
Inserting Eq.~(\ref{nfold-pow-gen}) into Eq.~(\ref{epsilon-pow-gen}),
we get
\be
\epsilon_* \sim \frac{1}{4 (N_* + \tN)}\,.
\label{epsilon-pow-gen-n}
\ee
We see that for slow-roll inflation the slow-roll parameter $\epsilon$ is not influenced by cuscuton at the leading order if the cuscuton potential takes the quadratic form.
From Eq.~(\ref{slowroll-p}), the parameter $\alpha$ can be estimated as
\be
\alpha_* = \left. \frac{\dot \phi^2}{2 \Mp^2 H^2}\right|_* 
= \frac 12 \(\frac{d \phi_*}{d N_*}\)^2 
\simeq \frac 12 \(\Mp^2 - \frac 32 \Lambda\)^2 \frac{U_\phi^2}{U^2}\,
\sim \(\Mp^2 - \frac 32 \Lambda\) \frac{1}{4 \(N_* + \tN\)}\,,
\label{alpha-n-gen}
\ee
and hence the parameter $\beta$ is
\be
\beta_* = - \frac{1}{\alpha_*}\frac{d \alpha_*}{d N_*}
\sim  \frac{1}{N_* + \tN}\,.
\label{beta-n-gen}
\ee
It follows from the above estimations and Eq.~(\ref{predict}) that $n_s$ is not affected by the cuscuton at the leading order,
while $r$ can be suppressed.

We now consider the specific form of the inflaton potential.
Inserting Eqs.~(\ref{h2-pow}) and (\ref{inf-pot}) into Eq.~(\ref{e-folding}),
the number of e-folding can be computed as
\be
N_* = \int_{\phi_e}^{\phi_*}  \frac{1}{2\(\Mp^2 - 3 \Lambda / 2\)} \phi d \phi\,,
\label{e-folding-pow}
\ee
Applying the slow-roll approximation to Eq.~(\ref{h2-pow}) and differentiating the result with respect to time,
we get 
\be
\epsilon=  2 \(\Mp^2 - \frac{3}{2} \Lambda\) \frac{1}{\phi^2}\,.
\label{ep-pow}
\ee
At the end of inflation, $\epsilon = 1$ so that the above equation yields
\be
\phi_e^2 =  2 \(\Mp^2 - \frac{3}{2} \Lambda \)\,.
\label{end-pow}
\ee
This equation shows that $\phi_e$ is suppressed in cuscuton inflation compared with the standard one.
Substituting  this result into Eq.~(\ref{e-folding-pow}),
Eq.~(\ref{e-folding-pow}) gives relation between the inflaton field and the number of e-folding as
\be
\phi_*^2 = 4\(\Mp^2 - \frac 32 \Lambda \)\(N_* +\frac{1}{2}\)\,.
\label{phi*-pow}
\ee
Hence, Eq.~(\ref{ep-pow}) gives
\be
\epsilon_* =  \frac{1}{2 N_* + 1}\,.
\label{epn-pow}
\ee
The parameter $\alpha$ is computed by diffferentiating Eq.~(\ref{phi*-pow}) with respect to $N_*$ as
\be
\alpha_* = \frac{1}{2 \Mp^2}\(\frac{d\phi_* }{d N_*}\)^2
= \frac{1 - 3 \Lambda / (2 \Mp^2)}{2 N_* + 1}\,.
\label{al-pow}
\ee
From the above expression for $\alpha_*$,
we obtain
\be
\beta_* = \frac{2}{2 N_* + 1}\,.
\ee
Hence, the  scalar spectral index and the tensor-to-scalar ratio at the horizon crossing are
\be
n_s -1 = - \frac{4}{2 N_* + 1}\,,
\quad\mbox{and}\quad
r = \frac{16 - 24 \Lambda / \Mp^2}{2 N_* + 1}\,,
\label{nr-pow}
\ee
The observational bound on these parameters from Planck \cite{Planck:2018jri} are
\begin{align}
    n_s \sim 0.96\,, \qquad 
r < 0.1\,.
    \label{eq:Planck_constraints}
\end{align}
Setting $n_s$ in Eq.~(\ref{nr-pow}) according to the above observational value,
and plugging the obtained $N_*$ in the expression of $r$,
we get $r < 0.1$ if $\Lambda / \Mp^2 > 0.25$.
To avoid negative $r$, we require $\Lambda / \Mp^2 < 2 / 3$.

After the end of inflation, the inflaton oscillates around the minimum of its potential.
The oscillating solution can be estimated by using variable $\phi =a^{-3/2}Y$,
so that Eq.~(\ref{kgphi}) becomes
\begin{equation}
\ddot{Y} +\left(m_\phi^{2}-\frac{9}{4}H^{2}-\frac{3}{2}\dot{H}\right)Y = 0\,.
\label{Ybar Eq}
\end{equation}
According to the approximations $m_\phi^{2}\gg H^{2}$ and $m_\phi^{2}\gg|\dot{H}|$ after inflation,
Eq.~(\ref{Ybar Eq})  is satisfied by the solution in the form
\begin{equation}
Y = A \sin(m_\phi t)\,,
\end{equation}
where $A$ is a constant.
Hence, the solution of $\phi$ is
\begin{equation}
\phi = \frac{A}{a^{3/2}} \sin(m_\phi t)\,.
\label{phi_sol_a}
\end{equation}

It is worth mentioning here that a massive minimally coupled scalar field behaves like dust-like matter after inflation ($mt\gg 1$), so that $a(t)\propto t^{2/3}$ after averaging over time interval much larger than $m^{-1}$, was derived analytically in \cite{Starobinskii:1978}. Averaging over several oscillations of the inflaton, see also Ref.\cite{GarciaBellido:2008ab}, the energy and pressure densities associated to the solution (\ref{phi_sol_a}) are given by
\begin{eqnarray} \label{omega}
P_{\phi} \approx \left<\frac{1}{2}{\dot \phi}^{2} - \frac{1}{2}m^{2}_{\phi}\phi^{2}\right> 
\approx 0\,,
\end{eqnarray}
where we suppose $m_\phi > H$ after inflation.
This implies that the averaged energy density  of the inflaton behaves like matter,
and therefore Eq.~(\ref{h2-pow}) gives $a \propto t^{2/3}$.
As a result, Eq.~(\ref{phi_sol_a}) becomes
\begin{eqnarray} \label{sophif}
\phi(t) \approx \frac{\phi_{e}}{m_{\phi}t}\,\sin(m_{\phi}t)\equiv \Phi(t) \sin(m_{\phi}t)\,,
\label{phi-osc-pow}
\end{eqnarray}
where the amplitude $A$ of the oscillation is computed by matching $\phi(t \to 0) = \phi_e$ and $\phi_e$ in this case is given by Eq.~(\ref{end-pow}).
It follows from Eqs.~(\ref{end-pow}) and (\ref{nr-pow}) that the amplitude of inflaton oscillation decreases if $r$ decreases.

\subsection{The exponential potential}\label{ep}

We now consider the exponential potential in the form
\be
V(\varphi) = V_0 e^{-\kappa \varphi}\,,
\label{v-exp}
\ee
where $V_0$ and $\kappa$ are constant with dimension of mass${}^4$ and mas${}^{-1}$.
For the exponential potential,
Eq.~(\ref{kgcus}) yields
\be
    \mathrm{sign}(\dot \varphi) 3 \mu^2 H = \kappa V\,. 
\label{vh-exp}
\ee
According to Eqs.~(\ref{v-exp}) and (\ref{vh-exp}),
if we deman that the Hubble parameter is positive,
the field $\varphi$ has to increase with time for a positive $\kappa$ and has to decrease with time for a negative $\kappa$.
Hence, we can drop $    \mathrm{sign}(\dot \varphi)$ and suppose that $\kappa > 0$ in the following calculations.
The relation in Eq.~(\ref{vh-exp}) can be used to eliminate $V$ from Eq.~(\ref{h2}),
so that the friedmann equation becomes
\be
3 \(\Mp^2 H^2 - \frac{\mu^2}{\kappa} H\) = U + \frac{1}{2} \dot\phi^2\,. 
\label{h2-exp0}
\ee
Hence, the Hubble parameter is given by
\be
H = \Xi \pm \sqrt{\Xi^2 + \frac{1}{3\Mp^2} \(U + \frac{1}{2} \dot\phi^2\)}\,,
\label{h-sol}
\ee
where $\Xi \equiv \mu^2 / (2 \kappa \Mp^2)$.
In the following calculations we choose the solution with plus sign to avoid the negative $H$.
From Eq.~(\ref{h-sol}), we have
\be
3H^2 = 6\Xi^2 + \frac{\rho_\phi}{\Mp^2} + 2 \Xi \sqrt{9 \Xi^2 + \frac{3}{\Mp^2} \rho_\phi}\,,
\label{h2-exp-sol}
\ee
where $\rho_\phi \equiv \dot\phi^2 / 2 + U$ is the energy density of the inflaton.
To investigate situation in which the inflationary universe can be realized,
we compute the slow-roll parameter $\epsilon$ by differentiating Eq.~(\ref{h-sol}) with respect to time:
\be
\epsilon =
\frac{\sqrt{3}}{2 \Mp H} \frac{\dot\phi^2}{\sqrt{3 \Mp^2 \Xi^2 + \rho_\phi }}\,.
\label{ep-exp-gen}
\ee
The slow-roll parameter $\epsilon$ can be less than unity corresponding to inflationary phase and can be larger than one corresponding to epoch after inflation, i.e., the model has graceful exit,
if the energy density of inflaton $\rho_\phi$ is larger than $\Mp^2 \Xi^2$.
In this situation, the inflationary phase can be realized if the inflaton field slowly evolves.
Supposing that the energy density of inflaton is significantly larger than $\Mp^2 \Xi^2$,
Eq.~(\ref{h2-exp-sol}) can be approximated as
\be
3H^2 \simeq \frac{2\sqrt{3}}{\Mp} \sqrt{\rho_\phi} \Xi + \frac{\rho_\phi}{\Mp^2}\,.
\label{h2-approx}
\ee
Applying the slow-roll approximation to Eq.~(\ref{h2-approx}), we can write
\be
3H^2 \simeq \sqrt{6} \barm \phi \Xi + \frac{\barm^2}{2} \phi^2\,,
\label{h2-exp-ap}
\ee
where $\barm = m_\phi / \Mp$ and we have used the inflaton potential from Eq.~(\ref{inf-pot}).
Differentiating Eq.~(\ref{h2-exp-ap}) with respect to time,
the slow-roll parameter is obtained under our assumption as
\be
\epsilon = \frac{2 \Mp^2}{\phi} \frac{\sqrt{6} \barm \Xi + \barm^2 \phi}{\(2 \sqrt{6} \Xi + \barm \phi\)^2}\,.
\label{ep-exp-ap}
\ee
Setting $\epsilon = 1$,
the inflaton field at the end of inflation can be computed from the relation
\be
\phi_e \(2 \sqrt{6} \Xi + \barm \phi_e\)^2 = 2 \Mp^2 \(\sqrt{6} \barm \Xi + \barm^2 \phi_e\)\,.
\label{pe-exp}
\ee
If we suppose that $\phi \sim {\cal O}(\Mp)$,
the approximation $\rho_\phi \gg \Mp^2 \Xi^2$ is equivalent to $m_\phi \gg \Xi$.
Further assuming that $\barm < 1$,
the approximation becomes $\Mp \gg \Xi $.
The solutions for Eq.~(\ref{pe-exp}) under this approximation are
\be
\phi_e = \sqrt{2} \Mp - \frac{3 \sqrt{3/2}}{\barm} \Xi\,,
\quad
\phi_e = - \sqrt{2} \Mp - \frac{3 \sqrt{3/2}}{\barm} \Xi\,,
\quad
 \phi_e = - \frac{\sqrt{6}}{\barm}\Xi\,.
\label{phie-exp}
\ee 
We are interested in the first solution.
The above approximated $\phi_e$ differs by less than a few percent  from that numerically computedfrom Eq.~(\ref{ep-exp-gen}) under the slow-roll approximation if $\Xi / m_\phi \leq 0.05$.
Supstituting Eq.~(\ref{h2-exp-ap}) into Eq.~(\ref{e-folding}),
the number of e-folding for this case can be computed as
\be
N_* = \int_{\phi_e}^{\phi_*}  \frac{1}{m_\phi \Mp } \(\sqrt{6} \Xi + \frac{\barm}{2} \phi\) d \phi\,.
\label{e-folding-exp}
\ee
The above equation yields
\be
\phi_* = \frac{2}{\barm}\(- \sqrt{6} \Xi \pm \sqrt{6 \Xi^2 + m_\phi^2 \(N_* + \tN\)}\)\,,
\label{p*-n}
\ee
where 
\be
\tN \equiv \frac{4 \barm^2 \Mp^2 + 4 \sqrt{3} \barm \Mp \Xi - 45 \Xi^2}{8 m_\phi^2}\,.
\ee
We will consider the solution with the plus sign in the following analysis.
Substituting Eq.~(\ref{p*-n}) into Eq.~(\ref{ep-exp-ap}),
we get
\be
\epsilon_* = 
\frac{m_\phi^2}{4\(- \sqrt{6} \Xi + \sqrt{6 \Xi^2 + m_\phi^2 \(N_* + \tN\)}\)\sqrt{6 \Xi^2 + m_\phi^2 \(N_* + \tN\)}}
+ \frac{ m_\phi^2 }{4 \(6 \Xi^2 + m_\phi^2 \(N_* + \tN\)\)}\,.
\label{ep-exp-n}
\ee
The parameter $\alpha$ can be computed by differentiating Eq.~(\ref{p*-n}) with respect to $N_*$ as
\be
\alpha_* = \frac{m_\phi^2}{12 \Xi^2 + 2 m_\phi^2 \(N_* + \tN\)}\,.
\label{al-exp}
\ee
This expression of $\alpha_*$ yields
\be
\beta_* = \frac{m_\phi^2}{6 \Xi^2 + m_\phi^2 \(N_* + \tN\)}\,.
\ee
Hence, the predictions for this model are
\ba
n_s -1 &=& 
- \frac{m_\phi^2}{2\(- \sqrt{6} \Xi + \sqrt{6 \Xi^2 + m_\phi^2 \(N_* + \tN\)}\)\sqrt{6 \Xi^2 + m_\phi^2 \(N_* + \tN\)}}
- \frac{ 3 m_\phi^2 }{2 \(6 \Xi^2 + m_\phi^2 \(N_* + \tN\)\)}
\nonumber\\&\simeq&
- \frac{4}{2 N_*+ 1}
+ 2 \sqrt{3}\frac{\Xi}{m_\phi} \frac{2 - \sqrt{2 N_* + 1}}{\(2 N_* + 1\)^2}\,,
\\
r &=& 16 \frac{m_\phi^2}{12 \Xi^2 + 2 m_\phi^2 \(N_* + \tN\)}
\simeq \frac{16}{2 N_* + 1}
- \frac{16 \sqrt{3} \Xi}{m_\phi \(2 N_* + 1\)^2}\,.
\label{nr-exp}
\ea
We estimate how much  the observational quantities in this model of cuscuton inflation differ from those in standard inflation by computing the factions:
\ba
\frac{n_s - n_s^{(s)}}{n_s^{(s)} - 1} 
&\sim& -\frac{\sqrt{3}}{2\sqrt{2}}\frac{\Xi}{m_\phi \sqrt{N_*}} < - 0.01\,,
\label{ns-delta}\\
\frac{r - r^{(s)}}{r^{(s)}} &\sim& - \frac{\sqrt{3} \Xi}{2 m_\phi N_*}
< - 0.01\,,
\label{r-delta}
\ea
where superscript ${}^{(s)}$ denotes quantities from standard inflation, i.e., $\Xi = 0$.
The upper bound $\Xi / m_\phi < 0.05$ is used in the estimation of Eqs.~(\ref{ns-delta}) and (\ref{r-delta}) to ensure that the above approximations have error up to a few percent.

After inflation, the inflaton field starts to oscillate around local minimum of its potential.
To estimate the evolution of the scale factor,
we have to average Eq.~(\ref{h2-approx}) over several oscillations of the inflaton
\be
3 \langle H^2 \rangle \simeq \frac{2\sqrt{3}}{\Mp} \langle \sqrt{\rho_\phi}\rangle \Xi + \frac{\langle\rho_\phi\rangle}{\Mp^2}\,.
\label{h2-approx-avg}
\ee
According to Eq.~(\ref{kgphi}), the oscillation of inflaton depends on the form of its potential while the Hubble parameter plays a role of the dampping term.
Hence, the averaged evolution of inflaton for this case is the same as that in the previous section such that averaged energy density of inflaton behaves like matter.
According to Eq.~(\ref{phi_sol_a}),
we suppose that the damped oscillation of the inflaton can be described by
\be
\phi(t) = \phi_0(t) \sin(m_{\phi}t)\,,
\label{phit-gen}
\ee
so that the energy density of inflaton is
\be
\rho_\phi = \frac 12 \dot{\phi}_0^2 \sin^2(m_{\phi}t) + \dot{\phi}_0 m_\phi \phi_0 \sin(m_{\phi}t)\cos(m_{\phi}t) + \frac 12 m_\phi^2 \phi_0^2\,.
\ee
Since $\phi_0(t)$ is a function of a scale factor,
one can suppose that $\dot{\phi}_0/ \phi_0 \sim {\cal O}(H)$.
Using $m_\phi > H$ during preheating, we can approximate
\be
\sqrt{\rho}_\phi \approx \sqrt\frac{1}{2} \dot{\phi}(t)_0\sin(m_{\phi}t)\cos(m_{\phi}t) + \sqrt{\frac 12} m_\phi \phi_0(t)\,.
\label{sqrtrho}
\ee
Averaging Eq.~(\ref{sqrtrho}) over several oscillations of the inflaton,
we get $\langle \sqrt{\rho}_\phi\rangle \approx \sqrt{1/2} m_\phi \phi_0(t)$ and consequently Eq.~(\ref{h2-approx-avg}) yields
\be
3 H^2 \approx \sqrt{6} \barm \Xi \phi_0(t) + \frac 12 \barm^2 \phi_0^2(t)\,.
\label{h2-approx-avg1}
\ee
Setting $\phi_0(t) = A / a^{(3/2)}$,
the above equation can be integrated as
\be
4 \frac{\sqrt{\sqrt{6} \barm \Xi A a^{3/2} + \barm^2 A^2 / 2}}{3\sqrt{2} \barm \Xi A} \approx t + C\,.
\ee
The integration constant $C$ is chosen such that the above equation reduces to $a \propto t^{2/3}$ when $\Xi \to 0$,
so that we get $C = 2 / (3 \Xi)$ and therefore
\ba
a^{3/2} &\approx &  \frac{3 \sqrt{6} \barm \Xi A}{16} t^2 + \frac{\sqrt{3} \barm A}{2 \sqrt{2}} t
\nonumber\\
&\approx & \frac{\sqrt{3} \barm A}{2 \sqrt{2}} t\(1 + \frac{3 \Xi}{4} t\)\,.
\ea
Inserting this result into Eq.~(\ref{phi_sol_a}),
we get
\begin{equation}
\phi \approx B \(1 + \frac{3 \Xi}{4} t\)^{-1} \frac{2 \sqrt{2}}{\sqrt{3} \barm t}\sin(m_\phi t)\,,
\label{phiexp-f}
\end{equation}
where $B$ is a constant.
Matching $\phi(m_\phi t \to 0) \to \phi_e$,
we get
\be
\phi \approx 
\phi_e \(1 + \frac{3 \Xi}{4} t\)^{-1} \frac{\sin(m_\phi t)}{m_\phi t} \equiv \bar{\Phi}(t) \sin(m_\phi t)\,,
\label{phi-osc-exp}
\ee
where $\phi_e$ is given by Eq.~(\ref{phie-exp}).
After the end of inflation,
the averaged energy density of the inflaton evolves as $\rho_\phi \propto a^{-3}$.
Hence, the energy density of the inflaton can drop below $\Mp^2 \Xi^2$.
Consequently, the universe could start an inflationary phase again as followed from Eq.~(\ref{ep-exp-gen}).
This implies that $\Xi$ has to be suppressed as the universe evolves which could be achieved if $\mu$ or $\kappa$ depend on time,
otherwise $\Xi$  has to be small such that it has no effect during inflationary phase.

\section{Preheating}
\label{sec4}

In this section, we study parametric resonances of models when an inflaton field $\phi$ coupled to another scalar field $\chi$ with the interaction term $g^{2}\phi^{2}\chi^{2}$, 
so that the action for our model is
\ba
S &=& \int d^4x\,\sqrt{-g} \Bigg(
\frac{\Mp^2}{2} R + \mu^2 \sqrt{- \partial_\mu \varphi \partial^\mu \varphi} - V(\varphi) 
- \frac{1}{2} \partial_\mu \phi \partial^\mu \phi - U(\phi)
\nonumber\\&&\quad\quad\quad\quad\quad\quad -\frac{1}{2} \partial_\mu \chi \partial^\mu \chi - W(\chi) -\frac{1}{2} g^{2}\phi^{2}\chi^{2}
\Bigg)\,,\label{ac}
\ea
where $W(\chi)$ is the potential of the  field $\chi$ and $g$ is the coupling constant. In this case scenario, we will choose the potential $W(\chi)$ of the form:
\begin{eqnarray}
W(\chi) = \frac{1}{2}m^{2}_{\chi}\chi^{2} \,,
\label{consoE}
\end{eqnarray}
where the mass $m_{\chi}$ is constant. Models with different form of $W(\chi)$ may also receive physical interest, e.g., Ref.\cite{Kofman:1994rk,Kofman:1997yn} for a non-minimally coupled scenario of the form $\sim \xi R\chi^{2}$. The time evolution of the quantum fluctuations of the field $\chi$ is ruled by the classical equation of motion, a.k.a., the Klein-Gordan equation in an expanding flat FLRW universe. From Eq.~(\ref{ac}), it is rather straightforward to derive the equation of motion for the field $\chi$ to obtain
\begin{eqnarray}
\ddot{\chi}+3H\dot{\chi} - \frac{1}{a^{2}}\nabla^{2}\chi +\Big[m^{2}_{\chi} + g^{2}\phi^{2}\Big] \chi = 0\,,  \label{ST_2.2}
\end{eqnarray}
where an effective $\chi$ mass $m^{2}_{\rm eff.}=m^{2}_{\chi}+g^{2}\phi^{2}$. We then expand the scalar fields $\chi$ in terms of the Heisenberg representation to yield
\begin{eqnarray}
\chi(t,{\bf x}) \sim \int \left(a_{k}\chi_{k}(t)e^{-i{\bf k}\cdot{\bf x}} + a^{\dagger}_{k}\chi^{*}_{k}(t)e^{i{\bf k}\cdot{\bf x}}\right)d^{3}{\bf k} \,, \label{mospace}
\end{eqnarray}
where $a_{k}$ and $a^{\dagger}_{k}$ are annihilation and creation operators. We can show that $\chi_{k}$ obeys the following equation of motion:
\begin{eqnarray}
\ddot{\chi}_{k}+3H\dot{\chi}_{k} +\Big[ \frac{k^{2}}{a^{2}} + m^{2}_{\chi} + g^{2}\phi^{2}\Big]\chi_{k} = 0\,. \label{Pot211}
\end{eqnarray}
Performing Fourier transformation to this equation and rescaling the field using $Y_{k} = a^{3/2}\chi_{k}$, we have
\begin{eqnarray}
\ddot{Y}_{k} + \omega^{2}_{k}Y_{k} = 0\,, \label{Pot2111}
\end{eqnarray}
where a time dependent frequency of $Y_{k}$ is given by
\begin{eqnarray}
\omega^{2}_{k} = \frac{k^{2}}{a^{2}} + m^{2}_{\chi}+ g^{2}\phi^{2}\,. \label{Pot20}
\end{eqnarray}
As is expected, Eq.~(\ref{Pot2111}) describes an oscillator with a periodically changing frequency $\omega^{2}_{k} = \frac{k^{2}}{a^{2}} + m^{2}_{\chi}+ g^{2}\phi^{2}$. The physical momentum ${\bf p}$ coincides with ${\bf k}$ for Minkowski space such that $k=\sqrt{{\bf k}^{2}}$. The periodicity of Eq.~(\ref{Pot2111}) may drive the parametric resonance for modes with certain
values of $k$. We suppose that the effective mass of the field $\chi$ vanishes $m_{\chi}=0$ and neglect for a moment the expansion of the universe, taking $a=1$ in Eq.~(\ref{Pot211}).

Let us consider for the first scenario the quadratic cuscuton potential examined in subsection (\ref{qp}). The simplest way to describe this important effect is to make a change of variables $m_{\phi} t \equiv z$. This simplifies Eq.~(\ref{Pot2111}) to the well-known Mathieu equation governed by
\begin{eqnarray}
\frac{d^{2}Y_{k}}{dz^{2}} + \left(A_{k} - 2q\cos(2z)\right)Y_{k} = 0\,. \label{Mathieu1}
\end{eqnarray}
The frequency terms are 
\begin{eqnarray}
A_{k} = \frac{k^{2}}{m_{\phi}^{2}}+2q\,,\,\, q = \frac{g^{2}\Phi^{2}(t)}{4m_{\phi}^{2}} \,. \label{Matheiu2}
\end{eqnarray}
An important feature of the solution of Mathieu equation is the existence of an exponential instability $Y_{k} \propto \exp(\mu_{k}z)$. This instability corresponds to an exponential growth of occupation number of quantum fluctuation $n_{k}(t) \propto \exp(2\mu_{k}z)$ \cite{Kofman:1994rk,Kofman:1997yn}. The modes $Y_{k}$ with momentum corresponding to the center of the resonance at $k\sim m_{\phi}$ grow as $e^{qz/2}$ which in this work equals $e^{qz/2}\sim e^{g^{2}\Phi^{2}t/ 8m_{\phi}}$.
Then the number of $\chi$-particles grows as $e^{2\mu_{k}z}\sim e^{qz}\sim e^{g^{2}\Phi^{2}t/4m_{\phi}}$ \cite{Kofman:1994rk,Kofman:1997yn}. 
From Eq.~(\ref{phi-osc-pow}), we have $q= g^{2}\phi_e^{2}/4 t^2 m^{4}_{\phi}$, so that $q$ can be very large if $m_\phi \ll \Mp$.  
In this regime, the resonance occurs for a broad range of values of $k$. Therefore, the parameter $\mu_{k}$ can be also large in which the resonance occurs for modes with $k^{2}/m^{2}_{\phi}=A-2q$. Specifically, the broad resonance occurs above the line $A=2q$ on the stability/instability chart for the Mathieu equation. We recommend the readers to follow the work done by \cite{Kofman:1994rk,Kofman:1997yn}. for detailed analysis on the stability/instability chart for the Mathieu equation \cite{McLachlan1947}. 

In the second case in which the cuscuton potential takes the exponential form discussed in Sec.~(\ref{ep}), we can also have an equation which describes an oscillator with a periodically changing frequency $\omega^{2}(t) = k^{2} + g^{2}\bar{\Phi}^{2} \sin^{2} (m_{\phi}t)$. One can have a Mathieu equation with $A_{k} = \frac{k^{2}}{m_{\phi}^{2}}+2q$, where 
$q = \frac{g^{2}\bar{\Phi}^{2}}{4m_{\phi}^{2}}$. 
Similar to the case of the  quadratic potential, a broad parametric resonance for the field $\chi$ in Minkowski space for $q \gg 1$ in this scenario can be obtained if $m_\phi \ll \Mp$. However, $q$ decreases faster than the case of the quadratic  potential due to a factor 
$f \equiv (1 + 3 \Xi t/ 4)^{-1}$. 
For a rough estimation, $H \sim 1/t$ so that 
$3 \Xi t/ 4 \sim 3 \Xi / (4 H) \sim 
(3/4) (\Xi /m_\phi) (m_\phi/ H)$
 which can be in order of unity. This suggests that a factor $f$ can significantly alter the decreasing rate of $q$. To investigate how does the factor $f$ influence the broad parametric resonance,
we compute $q$ for two cases of potentials.
For the quadratic  potential,
we have
\be
q = \frac{g^{2}\Mp^{2}}{2m_{\phi}^4 t^{2}}\(1 - \frac{3}{2} \frac{\Lambda}{\Mp^{2}} \)^{2}\,,
\ee
while for the exponential potential, we have
\be
q = \frac{g^{2}\Mp^{2}}{2 m_{\phi}^{4} t^{2}\Big(1-\frac{3\Xi t}{4}\Big)^{2}} \Big(1+\frac{3\sqrt{3}\Xi}{2m_{\phi}}\Big)^{2}\,.
\ee
If we suppose that the broad resonance stops at $q \gtrsim 1$,
the broad resonance stops at
\be
t_{s\,q} \sim \frac{g \Mp}{\sqrt{2} m^{2}_{\phi}}\,,
\quad
t_{s\,e} \sim \frac{2}{3 \Xi}\(- 1 + \sqrt{1 + 3 \Xi t_{s\,q}}\)=\frac{2\,t_{s\,q}}{3 \Xi t_{s\,q}}\(- 1 + \sqrt{1 + 3 \Xi t_{s\,q}}\)\,,
\ee
for the quadratic ($t_{s\,q}$) and exponential ($t_{s\,e}$) potentials, respectively. For the exponential potential, when $3\Xi t_{s\,q}< 1$, we can expand $\sqrt{1 + 3 \Xi t_{s\,q}}\sim 1+3 \Xi t_{s\,q}/2$ to simply find that $t_{s\,e}=t_{s\,q}$. However, in case of $3\Xi t_{s\,q}> 1$, we instead find that $t_{s\,e}<t_{s\,q}$. The later case implies that the broad resonance in the exponential potential stops earlier than that of the quadratic one.
   
  Nevertheless, a sufficient broad parametric resonance could be achieved if $q$ is initially large.
Consider the case where the expansion of space is taken into account. Let's take only the first model and we see that the equation of motion for the scalar matter fields, Eq.~(\ref{Pot2111}), can still be written in the form of a simple harmonic oscillator with a time varying frequency. Using the the same technique of performing Fourier transformation and taking $Y_{k}(t) = a(t)^{3/2}\chi_{k}(t)$, we have
\begin{eqnarray}
\ddot{Y}_{k} + \omega^{2}(k,t)\,Y_{k} = 0\,, \label{Pot2111e}
\end{eqnarray}
where a time dependent frequency of $Y_{k}(t)$ in this case is given by
\begin{eqnarray}
\omega^{2}(k,t) = \frac{k^{2}}{a^{2}} + m^{2}_{\chi}+ g^{2}\phi^{2}-\frac{9}{4}H^{2}-\frac{3}{2}{\dot H}\,. \label{Pot20e}
\end{eqnarray}
It is clear that the above equation is not the Hill's equation anymore. To quantify the evolution of particular comoving modes, the effects of the space expansion on particle production can be traced by considering the flow lines to the Floquet chart \cite{Zlatev:1997vd}. Note that in the matter-dominated background with $U(\phi)= m_{\phi}^{2} \phi^{2}/2$, and $\Phi\propto a^{-3/2},\,k\propto a^{-1}$ and $3H^{2}=-2{\dot H}$, the last two terms cancel. A given comoving mode $k$ flows along $\Phi\sim k^{3/2}_{phys}\equiv (k/a)^{3/2}$ curve in $k_{phys}-\Phi$ plane, see for example \cite{Zlatev:1997vd}.

The parameter $q = g^{2}\Phi^{2}/4m^{2}_{\phi}$ depends on time via $\Phi\propto a^{-3/2}$. Additionally, we can check in which a resonance band in our process develops. Following \cite{Kofman:1997yn}, the number of the band in the theory of the Mathieu equation is given by $n=\sqrt{A}$. In our case, reheating occurs for $A\sim 2q$, i.e. $n\sim \sqrt{2q}\sim \sqrt{g^{2}\Phi^{2}/2m^{2}_{\phi}}$. Suppose we have an inflationary theory with $m_{\phi}\sim 10^{-6} \Mp$, and let us take as an example $g\sim 10^{-2}$. Then after the first oscillation of the field, we have $\Phi(t)\sim 0.2 \Mp \sqrt{\(1 - \frac{3}{2} \frac{\Lambda}{\Mp^2}\)}\sim 0.2 \Mp$ for $\Lambda \ll \Mp$, which corresponds to $q\sim 10^{6}$. This gives the band number $n \sim 1414$.

\section{Conclusions}
\label{sec5}

In this work, we investigate observational predictions and preheating in cuscuton inflation.
For the case where the potential of the cuscuton  takes the quadratic form,
the scalar spectral index is the same as for the standard inflationary model at the leading order if the slow-roll approximation is applied.
Under the slow-roll approximation, the tensor-to-scalar ratio can be reduced such that its value agrees with the Planck data for any form of inflaton potential.
If the cuscuton potential has exponential form,
the energy density of the inflaton is required to be larger than the energy scale $\Mp^2 \Xi^2$ to have a graceful exit.
According to this condition, the scalar spectral index and tensor-to-scalar ratio from this model deviate by a few percent compared with those from standard inflation. 
The existence of the constant energy scale $\Xi$ leads to the problem when the energy density of the inflaton drops below this energy scale after inflation because the universe could accelerate again.
This problem could be alleviated if $\Xi$ can vary with time or its value is small such that it has no effect during inflation but could drive cosmic acceleration at late time. 

We have also examined the particle production due to parametric resonances in both models. We demonstrate that in Minkowski space the stage of parametric resonances can be described by the Mathieu equation. For the case of quadratic potential, the amplitude of the driving force in the Mathieu equation has a similar form as that in the standard inflationary framework. Nevertheless, in the case of exponential potential,
the amplitude of the driving force decreases faster than in the standard case. We find that the parametric resonances in our models can be sufficiently broad possible for the exponential growth of the number of particles. We also discuss the case in which the expansion of space is considered.

\section*{Acknowledgments}

P. Channuie acknowledged the Mid-Career Research Grant 2020 from National Research Council of Thailand (NRCT5-RSA63019-03). This work is also financially supported by the Thailand National Science, Research and Innovation Fund (TSRF) via PMU-B with grant No.R2565B030.

\end{document}